\documentstyle[mprocl]{article}

\def\be{\begin{equation}}
\def\ee{\end{equation}}
\def\bea{\begin{eqnarray}}
\def\eea{\end{eqnarray}}
\begin{document}

\title{Alternative Gravitational Theories in Four
  Dimensions\footnote{Report of parallel session chair in: Proc. 8th
    M. Grossmann Meeting, T. Piran (ed.) World Scientific, Singapore
    1998, to be published}}

\author{Friedrich W. Hehl}
\address{Institute for Theoretical Physics,
  University of Cologne, D-50923 K\"oln, Germany\\
email: \texttt{hehl@thp.uni-koeln.de}
}

\maketitle

\abstracts{We argue that from the point of view of gauge
  theory and of an appropriate interpretation of the interferometer
  experiments with matter waves in a gravitational field, the
  Einstein-Cartan theory is the best theory of gravity available.
  Alternative viable theories are general relativity and a certain
  teleparallelism model. Objections of Ohanian \& Ruffini against the
  Einstein-Cartan theory are discussed. Subsequently we list the
  papers which were read at the `Alternative 4D Session' and try to
  order them, at least partially, in the light of the structures
  discussed. }
  
\section{The best alternative theory?}

I would call general relativity theory$\,$\cite{meaning} GR the best
available \emph{alternative} gravitational theory and the next best
one its teleparallel equivalent$\,$\cite{Eggannalen,PRs} GR$_{||}$.
Because of these two theories, at least, it is good to have this
alternative session during the Marcel Grossmann Meeting. Let me try to
explain why I grant to GR the distinction of being the best alternative
theory.

After finally having set up special relativity theory in 1905,
Einstein subsequently addressed the question of how to generalize
Newton's gravitational theory as to make it consistent with special
relativity, that is, how to reformulate it in a Poincar\'e covariant
way. Newton's theory was a battle tested theory in the realm of our
planetary system and under normal laboratory conditions. It has
predictive power as it had been shown by the prediction of the
existence of the planet Neptune in the last century. The planets are
considered as point particles in this context, and they move in the
central gravitational field of the Sun. The attraction of neighboring
planets are accounted for by a highly developed perturbation theory.
Only very small deviation from the predictions of Newton's theory
puzzled a few experts by the end of last century.  But, for reasons of
{\em consistency}, Einstein had no other choice than to `marry'
Newton's gravitational theory to special relativity.

We all know the outcome of this undertaking: Special relativity turned
out to be too narrow.  Because of the equivalence principle, it had to
be a {\em curved} spacetime where gravity is appropriately housed. And
the simplest Lagrangian proportional to the curvature yields the left
hand side of Einstein's field equation which, in turn, explains the
post-Newtonian pieces of the perihelion advances of the planets.

The typical building blocks of GR are as \emph{sources} of gravity
fluid matter (in fact, usually ideal Euler fluids, without viscosity)
and the electromagnetic field, and as \emph{test `bodies'} classical
(structureless) point particles and light rays. Clearly, this picture
can be refined, but it is basically this scenario which we meet in GR.
Such a refinement you can see at work, if the motion of a narrow
binary pulsar is followed up in a general-relativistic computation,
and if radiation reaction terms are calculated, e.g.

In a reconstruction of the Riemannian spacetime of GR by a so-called
axiomatic approach it is then only logical to take the paths of
(massive) point particles and radar signals (`light rays', `photons')
as basic notions which implicitly define, by means of the axioms, the
spacetime of GR. Needless to say that point particles don't exist in
GR (they always carry a finite however small Schwarzschild radius with
themselves) and that the light rays are only a result of the
geometric optics limit in the framework of electrodynamics, i.e., a
short wavelength or high energy approximation of electrodynamics (see,
Mashhoon$\,$\cite{Bahram}). Therefore, in our axiomatic tool box, we have
little black holes and high-energy $\gamma$-rays at our disposal. 

Accordingly, the consistency question posed above by Einstein had been
answered for gravitating fluids, electromagnetic fields, massive point
particles, light rays.  And Einstein himself found the answer with his
gravitational theory of 1915/ 16, that is, in the framework of GR.

This would be the end of the story, if not a \emph{new} consistency question
had turned up with...

\section{`The Dawning of Gauge Theory'}

When modern gauge theories were developed in the
fifties,$\,$\cite{Lochlainn} it was the \emph{matter field} $\Psi$,
first as classical field, afterwards as second quantized field
operator -- rather than a hydrodynamical model of matter by means of
an ideal fluid -- which was the basis for the description of our
material surrounding. The matter field constitutes the Lagrangian
whose conserved currents and symmetries were studied.

The gravitational field was `rewritten' in terms of tetrads and the
gauge symmetry investigated. This turned out to be more than a sheer
rewriting: It was recognized in this context that the underlying gauge
group of gravity is represented by the group of motion of
\emph{special} relativity, namely the Poincar\'e
group$\,$\cite{GronwaldHehl}. That only its translational subgroup
should yield GR can be seen by pure counting of the corresponding
gauge potentials and the sources coupled to them: The $4\oplus
6$-dimensional Poincar\'e group has as its attached independent
currents the $4\oplus 6$ three-forms $\Sigma_\alpha$ (energy-momentum)
and $\tau_{\alpha\beta}=-\tau_{\beta\alpha}$ (spin), respectively.

Nevertheless, conventional prejudice has it that GR is either a gauge
theory of the diffeomorphism group or of the Lorentz group.  And even
if Feynman explicitly declares himself$\,$\cite{Feyn} that ``gravity is
that field which corresponds to a gauge invariance with respect to
displacement transformations'', i.e., with respect to translations,
you can be sure that his modern interpreters (in the
foreword$\,$\cite{Feyn} on p.\ XI) turn this around into the statement that
``the requisite gauge principle can be shown to be general
covariance.'' A review of the gauge theoretical aspects of gravity
theory was given by Gronwald and us$\,$\cite{GronwaldHehl}.

To cut a long story short, the Poincar\'e gauge theory of gravity
leads to a Riemann-Cartan geometry of spacetime, with curvature
$R^{\alpha\beta}$ and torsion by $T^\alpha$, and its simplest
Lagrangian, the curvature scalar of the Riemann-Cartan spacetime, to
the field equations of the so-called Einstein-Cartan theory of
gravity. Thus, within the modern paradigm of the gauge principle, the
consistency question had to be rephrased: It is no longer the matter
fluid or the test particle around which the theory revolves, rather
the matter field $\Psi$ is at the center of the stage.

At this point one could argue, as most of the `general relativists' in
fact do, that one should not care what those gauge theoreticians did
to Einstein's beautiful gravitational theory and should concentrate on
working out GR. Well, such a decision is perfectly possible as long as
you close your eyes$\,$\cite{dInverno,OR} to some more recent
\emph{experiments} in gravitational physics.

\section{The COW and BW experiments}

{\em Neutrons, atoms, and molecules} are the smallest and `most
microscopic' objects which, if exposed to the gravitational field,
show measurable effects on their wave functions. In the
Colella-Overhauser-Werner$\,$\cite{COW} and the
Bonse-Wroblewski$\,$\cite{BW} neutron interferometer experiments a
gravitational or acceleration induced phase shift of the wave function
has been observed with an accuracy of a few percent. With much greater
accuracy, such phase shifts have been verified in the Kasevich-Chu
atom interferometer experiments, see refs.$\,$\cite{Borde,Young}.
This phase shift can be calculated by means of the Schr\"odinger
equation with an external Newtonian gravitational potential. However,
there also exists the Einsteinian procedure for discovering the
effects of gravity.

But beware, we should {\em not} put the neutron matter wave or the
atomic beam including their respective interferometers on top of the
prefabricated Riemannian spacetime of GR in order to find out about
the effect of gravity on them. This is not what Einstein taught us,
since GR was only constructed for point particles, ideal fluids,
electromagnetic fields, etc.  Rather we should put (in a Gedanken
experiment) the experimental set-up, including the neutron wave etc.,
in a special-relativistic surrounding (that is, in a region where we
can safely neglect the gravitational field) and should wonder what
happens if, from the initial inertial reference frame, we go over to a
non-inertial frame.  This is what the quoted gauge theoreticians did
(in 1956 and 1961) with a matter wave function before the COW
experiment had even been conceived (around 1974). Technically, one has
to study how the special-relativistic Dirac Lagrangian -- if we
describe the neutron wave function approximately by means of a Dirac
spinor -- responds to the introduction of non-inertial reference
frames, see ref.$\,$\cite{GronwaldHehl}.

Accordingly, we apply the old Einsteinian procedure to a somewhat more
refined object than Einstein did between 1907 and 1915, namely to the
matter wave $\Psi$. Is it then a big surprise that the spacetime
emerging from this `modernized' Einstein procedure is a spacetime with
\emph{post}-Riemannian structures, or, to be more precise, with a
Riemann-Cartan structure?  This had been foreseen by E.\ 
Cartan$\,$\cite{Cartan} and, for the reason that in these spacetimes
locally, in a suitable \emph{normal} frame, metric and connection look
Minkowskian, Cartan called them spacetimes with a [pseudo-] Euclidean
connection.

Therefore the gauge paradigm and the COW-type experiments as well
suggest the emergence of a Riemann-Cartan spacetime. The simplest
nontrivial Lagrangian yields the...

\section{Einstein-Cartan theory of gravity}

Let us first define, in terms of the coframe $\vartheta^\alpha$ and
the Hodge star $^\star$, the 1-form $\eta_{\alpha\beta\gamma} =
{}^\star(\vartheta_\alpha\wedge\vartheta_\beta\wedge
\vartheta_\gamma)$ and the 3-form
$\eta_\alpha={}^\star\vartheta^\alpha$.  Then the two field equations
of Einstein-Cartan theory read,
\begin{eqnarray} {1\over 2}\,
  \eta_{\alpha\beta\gamma}\wedge R^{\beta\gamma}+ \Lambda\,\eta_\alpha
  &=& \ell^2\, \Sigma_\alpha\,,\label{ECfirst} \\ {1\over 2}\,
  \eta_{\alpha\beta\gamma}\wedge T^\gamma \!\,\;\quad &=&
  \ell^2\,\tau_{\alpha\beta} \,,\label{ECsecond}\end{eqnarray} where
$\Lambda$ denotes the cosmological constant and $\ell^2$ Einstein's
gravitational constant. 

The Einstein-Cartan theory is a viable gravitational theory. All
experiments known are correctly predicted by the theory.  One should
add, however, that under usual conditions the spin
$\tau_{\alpha\beta}$ of matter can be neglected which, in turn,
according to the second field equation, yields vanishing torsion --
and then we fall back to GR and to its predictions. The
Einstein-Cartan theory, as compared to GR, carries an additional
spin-spin contact interaction of gravitational origin. This additional
interaction only shows up at extremely high matter densities ($\;\sim
10^{54}\!$ g/cm$^3$ for neutrons) and hasn't been seen so far. Remember
that even in neutron stars we have only densities of the order of
$10^{15}\!$ g/cm$^3$. And in vacuum, according to the Einstein-Cartan
theory, there is no torsion which is consistent with a recent finding
of L\"ammerzahl$\,$\cite{Lammerzahl}.

Nevertheless, from a theoretical point of view, the Einstein-Cartan
theory appears to be \emph{the} gravitational theory. And, no doubt,
GR is the best alternative. This is why I called in the first section
GR to be an alternative theory. Of course, it depends on your
`reference frame' what you are inclined to call `alternative
gravitational theories'. I believe that the organizers of the Eighth
Marcel Grossmann Meeting didn't want me to interpret GR as an
alternative theory. Quite the opposite, Hans Ohanian and Remo Ruffini
even claim that the Einstein-Cartan theory is defective, see
ref.\cite{OR}, pp.\ 311 and 312. Since this is a widely read and,
otherwise, excellent textbook, I would like to comment on their
arguments:

\begin{itemize} \item \emph{O\&R, pp.311--312 (`local flatness syndrome'):}

\emph{``If $\,\Gamma^\beta{}_{\nu\mu}$ were not symmetric, the parallelogram
would fail to close.  This would mean that the geometry of the curved
spacetime differs from a flat geometry even on a small scale -- the
curved spacetime would not be approximated locally by a flat
spacetime."}

If an orthonormal coframe $\vartheta^\alpha=e_i{}^\alpha\,dx^i$ and a
linear connection
$\Gamma^{\alpha\beta}=\Gamma_i{}^{\alpha\beta}\,dx^i$ are given as
gravitational potentials, then, by a suitable coordinate {\it and} a
frame transformation, it can be shown$\,$\cite{vdH1,hart,Iliev} that
at one point P of a Riemann-Cartan manifold these potentials can be
`normalized' according to
\begin{equation}\{\vartheta^\alpha,\,\Gamma^{\alpha\beta}\}= 
  \{\delta^\alpha_i\,dx^i,\,0\} \qquad{\rm at\> one\> point\>
    P}\,.\label{vdh}\end{equation}
This is the analog of the Einstein elevator.

The O\&R statement on the approximate local flatness is only correct,
if one restricts oneselves to a {\it coordinate} (or natural) frame.
But it is incorrect in general, see eq.(\ref{vdh}). This was already known to
Elie Cartan in 1923/24 and, as mentioned above, it was this reason
which gave him the idea to name Riemann-Cartan spaces as spaces with
Euclidean connection.

\item  \emph{O\&R, p.312 footnote (`shaky spin discussion'):}

  \emph{``...we do not know the `genuine' spin content of elementary
    particles...''}

According to present day wisdom, matter is built up from quarks and
leptons. No substructures have been found so far. According to the
mass-spin classification of the Poincar\'e group and the experimental
information of lepton and hadron collisions etc., leptons and quarks
turn out to be fermions with spin 1/2 (obeying the Pauli principle).
According to an appropriate interpretation of the Einstein-Cartan
theory, see ref.$\,$\cite{He5}, the spin (of the Lorentz subgroup of
the Poincar\'e group) represents the source of torsion. As long as we
accept the (local) Poincar\'e group as a decisive structure for
describing elementary particles, there can be no doubt what spin
really is. And abandoning the Poincar\'e group would result in an
overhaul of (locally valid) special relativity theory.

The nucleon is a composite particle and things related to the build-up
of its spin are not clear so far. But we do know that we can treat it
as a fermion with spin $1/2$. As long as this can be taken for
granted, at least in an effective sense, we know its spin and
therefore its torsion content.
\end{itemize}

\noindent Of course, whether a theory is correct, can only be verified (or
falsified) by experiment. But the two points of O\&R, since both
incorrect, are irrelevant for this question.--- \bigskip

Let us come back to Einstein-Cartan theory proper. Recognizing the
central position of the matter field $\Psi$, Audretsch \&
L\"ammerzahl$\,$\cite{AL,AHL} initiated a new axiomatics for spacetime
which is based on wave notions. 

From the point of view of a Riemann-Cartan space, you can end up with
a flat Minkowski space in two ways: You can either start with a
Riemann-Cartan space and equate the torsion to zero, as in eq.(2), if
matter spin vanishes, or you can first put the (Riemann-Cartan)
curvature to zero, then one arrives, after this first step, at a
teleparallel (or \emph{Weitzenb\"ock}) space. Torsion still exists therein
and gravitational teleparallelism models with quadratic torsion
Lagrangians can be developed. One model mimicks perfectly well GR; for
a detailed discussion one should compare Mielke$\,$\cite{Eggannalen}.

The emerging Riemannian or Weitzenb\"ock spaces are, in some sense,
equivalent to each other. In a second step, one then puts curvature or
torsion, respectively, to zero and eventually reaches the Minkowski
space on both ways.

\section{Sessions on alternative theories}

Besides GR proper, d'Inverno$\,$\cite{dInverno} lists
\emph{Alternative theories, Unified field theory, and Quantum
  gravity}. Under `Alternative theories' we find the entries torsion
theories, Brans-Dicke, Hoyle-Narlikar, Whitehead, bimetric theories,
etc., under `Unified field theory' only Kaluza-Klein, and under
`Quantum gravity' canonical gravity, quantum theory on curved
backgrounds, path-integral approach, supergravity, superstrings, etc.

The organizers of the present Marcel Grossmann Meeting, in setting up
the parallel sessions, must have had in mind such a division.
Originally the present `Alternative session' encompassed also papers
on higher and lower dimensions. But these papers were so numerous that
they later had to be shifted to a newly installed session chaired by
Professor V.N.\ Melnikov (Moscow). Therefore in this session only 4D
papers were read.

In chronological order, I will list all the lectures which actually
took place. Some of the authors who submitted abstracts and who were
supposed to present their material didn't come.  They are not listed,
unless their paper was read by somebody else. It is for that reason
that the different subsessions are of quite different length. In
organizing the session, I tried to order the papers logically
depending on the topic they treat.  This was not possible for the
post-deadline papers. For each session I selected three main
contributions which I believed to be of general interest.

\subsection{Session I on Monday}

\qquad$\>\>${\em Three main contributions}

\begin{enumerate}

\item C. Rovelli: General relativity in terms of Dirac eigenvalues

\item (a) R.M. Zalaletdinov: Averaging problem in general relativity,
macroscopic gravity and using Einstein's equations in cosmology

(b) M. Mars, R.M. Zalaletdinov: Space-time averages in macroscopic gravity
and volume-preserving coordinates 

\item S.R. Valluri, W.L. Harper, R. Biggs: Newton's precession theorem,
eccentric orbits and Mercury's orbit \bigskip

{\em PPN and Newton-Cartan theory}

\item E.E. Flanagan: Coordinate invariant formulation of post-1-Newtonian
general relativity \bigskip

{\em New mathematical structures}

\item R.M. Santilli: Isotopic grand unification with the inclusion of
gravity (read by F.W.\ Hehl)

\item A.Yu. Neronov: General relativity as continuum with
microstructure. Formal theory of Lie pseudogroups approach\bigskip

{\em Theories based on a Riemann-Cartan-Weyl geometry of spacetime}

\item R.T. Hammond, C. Gruver, P. Kelly: Scalar field from Dirac coupled
torsion\bigskip

{\em Frames, teleparallelism, bimetric theories}

\item S. Kaniel, Y. Itin: Gravity by Hodge de-Rham Laplacian on frames

\item V. Olkhov: On thermal properties of gravity \bigskip

{\em Electromagnetism and gravity}

\item R. Opher,U.F. Wichoski: On a theory for nonminimal
gravitational-electro\-mag\-netic coupling consistent with observational
data

\item C. L\"ammerzahl, R.A. Puntigam, F.W. Hehl: Can the
electromagnetic field couple to post-Riemannian structures?\bigskip

\noindent{\em Additional post-deadline papers}

\item C.M. Zhang: Gravitational spin effect on the magnetic
  inclination evolution of pulsars

\item B.G. Sidharth: Quantum mechanics and cosmology -- alternative
  perspective
\end{enumerate}

\subsection{Session II on Thursday}

\qquad$\>\>${\em Three main contributions}

\begin{enumerate}\setcounter{enumi}{13}

\item Dj. \v Sija\v cki: Towards hypergravity

\item A.Burinskii: Spinning particle as superblackhole

\item C. L\"ammerzahl: New constraints on space-time torsion from
  Hughes-Drever experiments \bigskip

{\em Theories based on a metric-affine geometry of
spacetime}

\item S. Casanova, G. Montani, R. Ruffini, R. Zalaletdinov: On the
non-Riemannian manifolds as framework for geometric unification
theories

\item (a) A.V. Minkevich: Some physical aspects of gauge approach to gravity

(b) A.V. Minkevich, A.S. Garkun: Some regular multicomponent isotropic
models in gauge theories of gravity

\item F.W. Hehl, Yu.N. Obukhov: Is a hadronic shear current one of the
sources in metric--affine gravity (MAG)? \bigskip

{\em Other theories}

\item L.V.Verozub: Gravity: Field and Curvature (read by F.W. Hehl)\bigskip

{\em Additional post-deadline papers}

\item D. Rapoport: Torsion and quantum and hydrodynamical fluctuations

\item L.C. Garcia de Andrade: On non-Riemannian domain walls 

\item Kong: On Einstein-Cartan cosmology

\end{enumerate}

\subsection{Some remarks to the papers}

With the titles of the subsessions, I tried to circumscribe the
content of the corresponding papers. Papers on \emph{Riemann-Cartan}
spacetimes, in particular Einstein-Cartan type theories, or slight
generalizations therefrom, can be found under the
numbers$\,$\footnote{These numbers refer to the papers listed in
  Sec.5. Please, don't mix them up with the references!} 7 (enriched
by an additional Weyl field), 12, 16, 21, 22, 23. There were three
papers on \emph{frame and teleparallelism} theories: Number 8, which
was published in the meantime$\,$\cite{kani97} [and which we found
particularly interesting, see ref.$\,$\cite{kaniel14}], number 9, and
number 20.
 
If one relaxes the metric compatibility of the connection
$\Gamma_\alpha{}^\beta$ of spacetime, then one has the nonmetricity
$Q_{\alpha\beta}:= -\stackrel{\Gamma}{D}g_{\alpha\beta}$ as an
additional structure to play with. Such theories are of the type of
metric-affine gravity MAG$\,$\cite{PRs,OVEH,colliding}. The lecture of
\v Sija\v cki falls under this general heading, even if he has a
sizeable additional input from particle physics; also the papers
number 11, 17, 18, and 19 use the metric-affine structure. These
papers exhaust the gauge type papers in the sessions.

GR (more or less) was the basis of the papers number 1 (Euclidean
signature), 2, 4, and 10.  A historical lecture on precession effects
within Newtonian gravity (and what we can learn from it for GR) can be
found under number 3.  The paper number 15 describes some
supersymmetric generalization of the Kerr solution of GR.

After all these `conventional' alternative attempts, three papers are
left in which new mathematical structures (papers number 5 and 6) and
new quantum physical attempts were investigated.

I apologize for all inappropriate classification attempts!

\section*{Acknowledgments} I would like to thank Remo Ruffini and Tsvi 
Piran for the invitation to chair the `Alternative 4D Parallel Session'.

%\section*{Appendix}

\end{document}